\documentclass[iicol,sn-mathphys-num]{sn-jnl}% referee option is meant for double line spacing

\usepackage[T1]{fontenc}
\usepackage{lmodern}
\usepackage{graphicx}%
\usepackage{multirow}%
\usepackage{amsmath,amssymb,amsfonts}%
\usepackage{amsthm}%
\usepackage{mathrsfs}%
\usepackage[title]{appendix}%
\usepackage{xcolor}%
\usepackage{textcomp}%
\usepackage{manyfoot}%
\usepackage{booktabs}%
\usepackage{algorithm}%
\usepackage{algorithmicx}%
\usepackage{algpseudocode}%
\usepackage{listings}%
\usepackage{subcaption}
\usepackage{tabularx}
\usepackage{array} 
\usepackage{adjustbox}
\theoremstyle{thmstyleone}%
\theoremstyle{thmstyletwo}%

\theoremstyle{thmstylethree}%

\begin{document}

\title[Effective Mass in Quantum Hadrodynamics-I and its Impact on the Equation of State of Neutron Matter]{Effective Mass in Quantum Hadrodynamics-I and its Impact on the Equation of State of Neutron Matter}


\author*{\fnm{Ghitha N.A.} \sur{Rahiemy}}\email{ghitha.nad2003@mail.ugm.ac.id}

\author{\fnm{Eko T.} \sur{Sulistyani}}\email{sulistyani@mail.ugm.ac.id}

\author{\fnm{Pekik} \sur{Nurwantoro}}\email{pekik@ugm.ac.id}

\affil{\orgdiv{Department of Physics}, \orgname{Universitas Gadjah Mada}, \orgaddress{\street{Jalan Sains, Bulaksumur}, \postcode{55281}, \state{Yogyakarta}, \country{Indonesia}}}

\abstract{
The effective nucleon mass $M^*$ plays a central role in Quantum Hadrodynamics-I (QHD-I), linking scalar meson interactions at the microscopic level to the macroscopic properties of dense nuclear matter. In this work, we re-derive the scalar density integral in detail and validate it numerically using Gaussian quadrature. The numerical and analytic results are found to be in excellent agreement, confirming the robustness of both approaches. We then investigate the sensitivity of $M^*$ to different parameter sets, highlighting its strong influence on nuclear saturation, compressibility, and the resulting equation of state (EoS). The analysis shows that variations in meson–nucleon couplings propagate directly into differences in pressure and energy density, affecting the stiffness of the EoS. While QHD-I produces characteristically stiff EoS, the effective mass evaluation provides a transparent framework for connecting microscopic meson dynamics to macroscopic neutron star properties. These findings underline the relevance of $M^*$ as a microscopic–macroscopic bridge and demonstrate the utility of numerical methods for extending relativistic mean-field models in nuclear astrophysics.
}




\maketitle

\section{Introduction}\label{sec1}

Understanding the behavior of strongly interacting matter at low temperature and extreme density is a central problem in nuclear and astrophysical physics. Neutron stars provide unique observational laboratories for such regimes, where matter is compressed well beyond nuclear saturation density and relativistic effects are essential \citep{antic,die}. Accurate theoretical descriptions of dense matter are therefore crucial for connecting microphysical interactions to macroscopic observables (masses, radii, tidal deformabilities) inferred from electromagnetic and gravitational-wave measurements.

Quantum Chromodynamics (QCD) is the fundamental theory of the strong interaction, but at nuclear energy scales the theory is strongly coupled and perturbative expansions break down. Direct QCD calculations of dense matter require nonperturbative techniques (e.g., lattice methods or functional approaches) that are computationally demanding and limited in applicability at high baryon density \citep{Frishman,dupuis,abbott,kumar}. For this reason, effective field theories that employ hadronic degrees of freedom provide a practical and physically motivated alternative for modeling bulk nuclear matter.

Quantum Hadrodynamics (QHD) is an effective relativistic framework in which nucleons interact via meson exchange. In its simplest realization, QHD-I (the Walecka model) retains only a scalar ($\sigma$) and a vector ($\omega$) meson; the scalar field induces attraction and reduces the nucleon effective mass, while the vector field provides short-range repulsion \citep{wala,sw,er,ho}. Within the relativistic mean-field (RMF) approximation these meson fields are replaced by their expectation values, yielding closed expressions for bulk quantities (baryon density, energy density, pressure) that depend on a self-consistent effective mass.

As observational and computational capabilities have improved, research has progressed beyond this minimal framework toward fully RMF, formulated as an extension of QHD by \citet{wala}. RMF has been widely applied in neutron star studies to incorporate additional physics, such as nonlinear scalar interactions, $\rho$ mesons, magnetic fields, and finite-temperature effects \citep{bednarek, huang, pais, scurto, wang}. This transition reflects not only the need to address empirical challenges—such as reproducing neutron star masses and radii—but also the feasibility of implementing more realistic models with modern computational tools, which reduce reliance on oversimplified assumptions.

The effective (Dirac) nucleon mass, $M^*$, is defined in natural units as \citep{wala}
\begin{equation}
	\label{eq:1.1}
	M^* \equiv M - g_s\phi_0,
\end{equation}
where $M$ is the free nucleon mass, $g_s$ the scalar coupling and $\phi_0$ the mean scalar field. In integral form the self-consistency condition for $M^*$ reads
\begin{align}
	\label{eq:1.2}
	M^* &= M - \frac{g_s^2}{m_s^2}\,\frac{\gamma}{(2\pi)^3}\int_{0}^{k_F}\! d^3k \;\frac{M^*}{\sqrt{k^2+M^{*2}}} \notag \\
	&= M - \frac{g_s^2}{m_s^2}\,\frac{\gamma}{2\pi^2}\int_{0}^{k_F}\! dk \; \frac{k^2 M^*}{\sqrt{k^2+M^{*2}}},
\end{align}
with $m_s$ the scalar meson mass, $\gamma$ the spin–isospin degeneracy and $k_F$ the Fermi momentum. \citet{wala} provided an analytic evaluation of the scalar-density integral,
\begin{align}
	\label{eq:1.3}
	\int_0^{k_F}\! dk\; \frac{k^2 M^*}{\sqrt{k^2+M^{*2}}} 
	= \frac{M^*}{2}\left[ k_F\sqrt{k_F^2+M^{*2}} \right. \notag \\
	- M^{*2}\ln\!\left.\left(\frac{k_F+\sqrt{k_F^2+M^{*2}}}{M^*}\right)\right],
\end{align}
which leads to the standard self-consistent equation used in QHD-I studies.

Although the analytic expressions are well known, a careful numerical re-evaluation of the scalar-density integral serves two purposes: (i) it verifies algebraic reductions and prefactors that are easily mistyped when reducing $d^3k$ integrals to one-dimensional form, and (ii) it provides a robust baseline and reusable numerical framework for applying the same machinery to extended RMF models that include additional mesons, nonlinearities, or finite-temperature and magnetic-field effects. In this work we evaluate the scalar-density integral using Gaussian quadrature, re-derive the analytic expression (Appendix \ref{secA1}), and use the resulting $M^*$ to compute the EoS for representative parameter sets. We then discuss how variations in $M^*$—controlled by meson couplings and masses—propagate to the pressure–energy density relation relevant for neutron star modeling and nuclear astrophysics applications.

\section{Numerical Methods}\label{sc:sec2}

To evaluate the effective mass and associated integrals in the QHD-I model, two numerical techniques were employed: the Secant method for solving the self-consistent effective mass equation and Gaussian quadrature for evaluating the scalar density integral. Both methods were implemented in a dedicated computational program developed for this study.

\subsection{Secant Method}

The effective mass $M^*$ is determined from the self-consistent condition (\ref{eq:1.2}), which can be recast as a root-finding problem,
\begin{align}
	\label{eq:2.1}
	f(M^*) &= M^* - M + \frac{g_s^2}{m_s^2}\frac{\gamma}{2\pi^2} \int_0^{k_F} dk \frac{k^2 M^*}{\sqrt{k^2+M^{*2}}} \notag \\
	&= 0.
\end{align}
Following \cite{wala}, $\gamma=2$ was used for pure neutron matter ($Z=0$) and $\gamma=4$ for symmetric nuclear matter ($N=Z$).  

The Secant method was chosen for solving $f(M^*)=0$ because it avoids the need for derivative evaluations required by Newton–Raphson, while maintaining rapid convergence for smoothly varying functions. Iterations were performed until successive estimates of $M^*$ differed by less than $10^{-3}$, with a maximum of 1000 iterations. To ensure robustness, additional checks were implemented: if the iteration produced a negative $M^*$, the value was reset to a small positive number ($M^*/M \approx 0.1$) to maintain physicality; and if convergence was not reached within the iteration limit, the algorithm flagged the solution as non-convergent.

\subsection{Gaussian Quadrature}

Numerical evaluation of the scalar density integral was performed using Gaussian quadrature with Legendre polynomials. The method approximates
\begin{equation}
	\int_a^b f(x)\,dx \;\approx\; \sum_{i=1}^N w_i f(x_i),
\end{equation}
where $x_i$ are the roots of the $N$-th Legendre polynomial and $w_i$ are the corresponding weights. A linear transformation maps the physical integration range $[0,k_F]$ to the standard interval $[-1,1]$.  

For this study, a 20-point quadrature was adopted. This order was chosen as a compromise: it provides convergence to the analytic result of equation (\ref{eq:1.3}) while avoiding unnecessary computational overhead from higher orders.

\subsection{Program Development and Validation}

\begin{figure}[!h] 
	\begin{center} 
		\includegraphics[width=\linewidth]{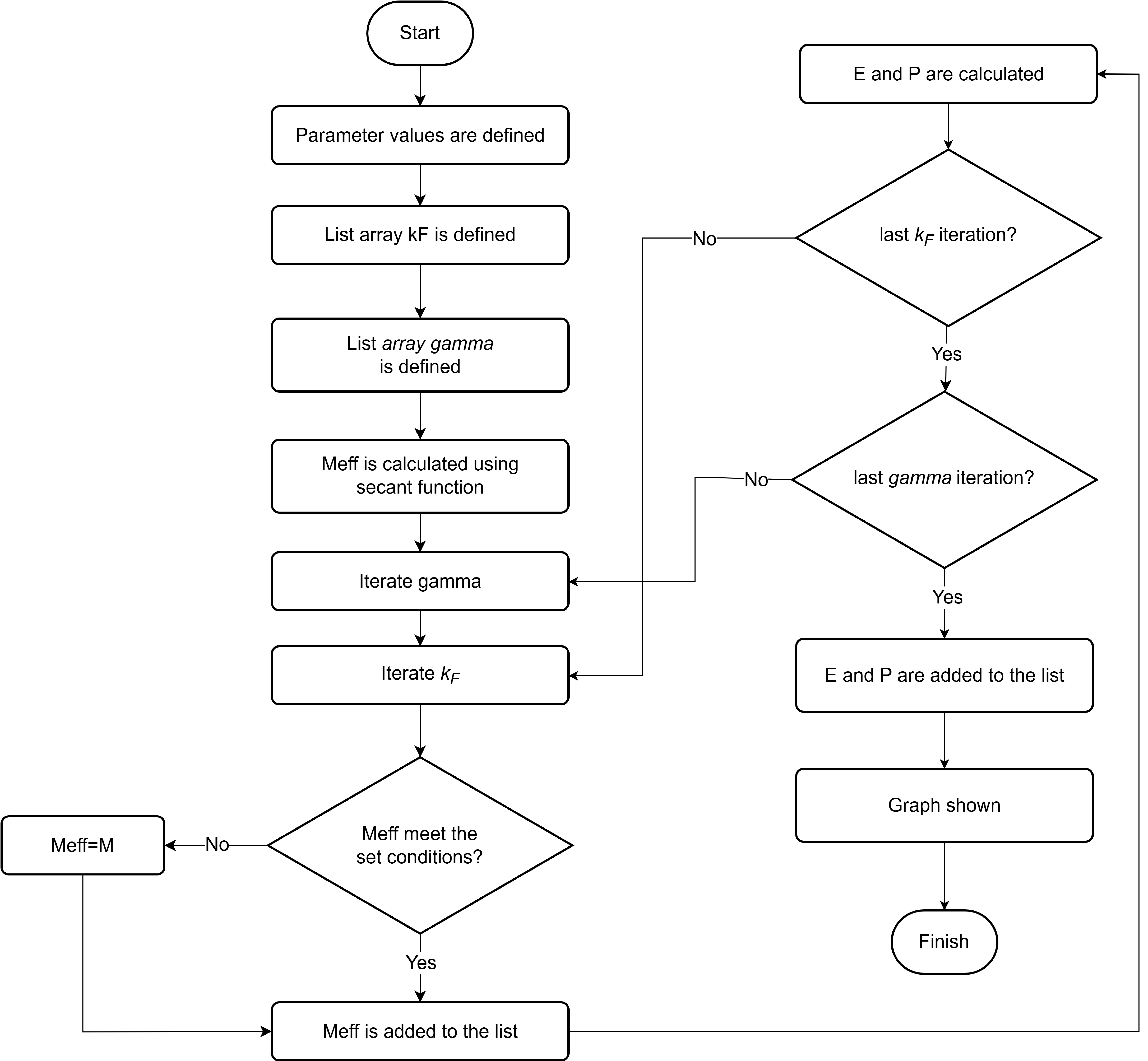} 
		\caption{Analytical evaluation program flowchart.} 
		\label{gb3.6} 
	\end{center} 
\end{figure}

\begin{figure}[!h] 
	\begin{center} 
		\includegraphics[width=\linewidth]{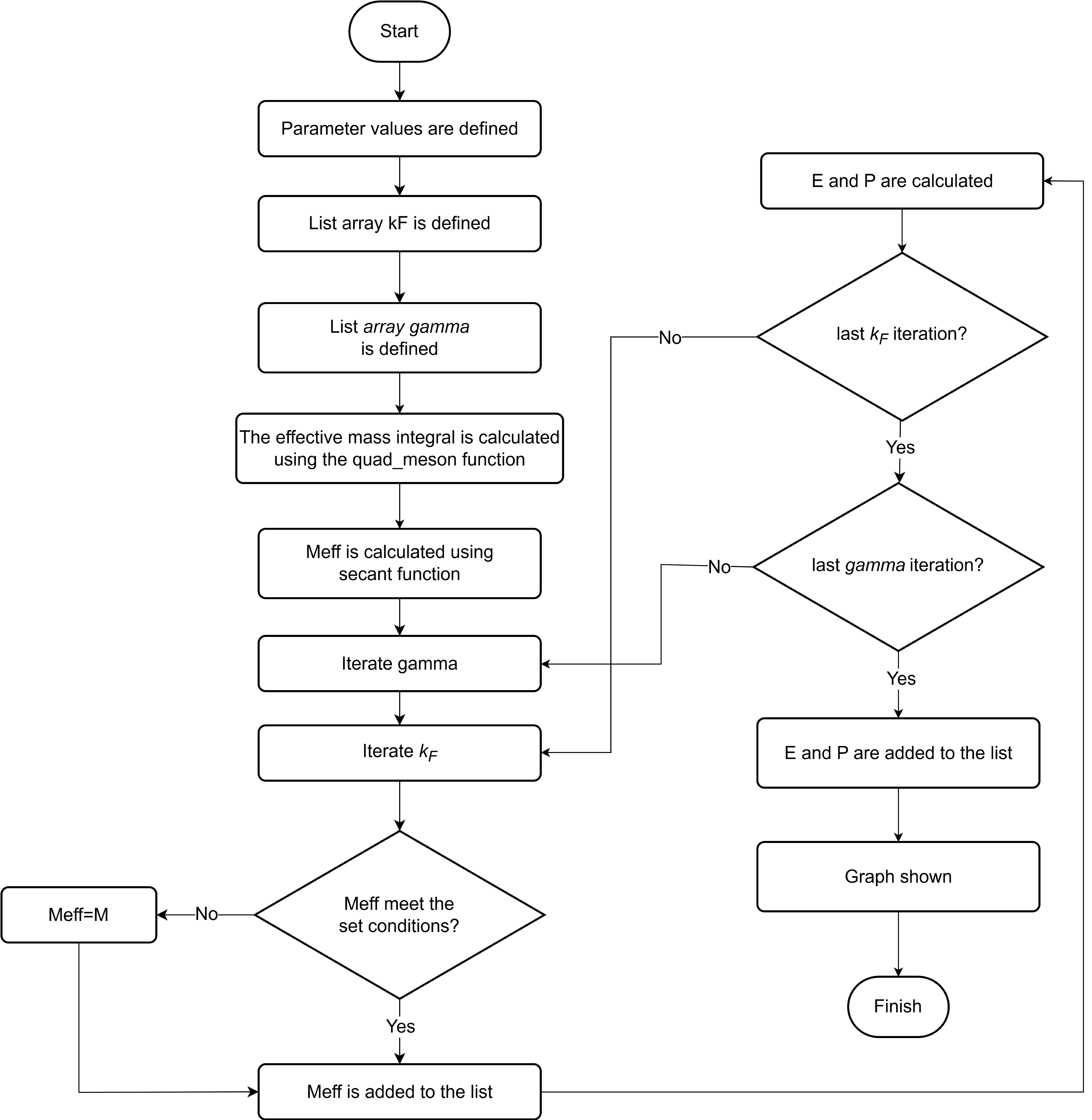} 
		\caption{Numerical evaluation program flowchart.} 
		\label{gb3.7} 
\end{center} \end{figure}

A computational program was developed to integrate these numerical methods into a unified framework. The workflow, summarized in Figures \ref{gb3.6} and \ref{gb3.7}, proceeds from parameter input (nucleon and meson masses, coupling constants) through evaluation of the scalar density integral and iterative solution of $M^*$. The program outputs include the effective mass, binding energy curves, and the pressure–energy density relation.

Validation was carried out in two steps: (i) the scalar density integral was computed using Gaussian quadrature and compared with the analytic expression to confirm accuracy, and (ii) the self-consistent solution for $M^*$ was benchmarked against published analytic results. Following validation, the program was applied to three parameter sets: \citet{wala}, \citet{swa}, and RHA0 \citep{swa}, summarized in Table \ref{tab:parameter_sets}. This setup enabled systematic exploration of how variations in meson masses and couplings influence the effective mass and derived nuclear matter properties.

\begin{table}[h!] 
	\centering 
	\caption{Parameter sets used in the numerical calculations of QHD-I. $m_s$ is the scalar meson mass, while $C_s^2 \equiv g_s^2 (\frac{M^2}{m_s^2})^2$ and $C_v^2 \equiv g_v^2 (\frac{M^2}{m_v^2})^2$. All sets use $M=939$ MeV and $m_v=783$ MeV.} 
	\label{tab:parameter_sets} 
	\begin{tabular}{c c c c} 
		\hline 
		\textbf{Parameter Set} & $m_s$ (MeV) & $C_s^2$ & $C_v^2$ \\ 
		\hline 
		\citet{wala} & 550.0 & 266.9 & 195.7 \\ 
		\citet{swa} & 550.0 & 357.4 & 273.8 \\ 
		RHA0 \cite{swa} & 456.0 & 228.0 & 147.5 \\ 
		\hline 
	\end{tabular} 
\end{table}

The workflow is also readily extendable to more sophisticated RMF models: replacing the scalar density integral with alternative expressions (e.g., nonlinear $\sigma$ interactions, $\rho$ mesons, or finite-temperature corrections) requires only minimal modification to the code. This flexibility allows systematic investigation of various RMF extensions and their effects on nuclear matter properties.

\section{Results and Discussion}

This section presents the numerical validation of the scalar density integral, the evaluation of the effective nucleon mass $M^*$ for different parameter sets, and the implications for the EoS of neutron matter in the QHD-I framework.

\subsection{Numerical Integration Validation}

\begin{figure}[t]
	\centering
	\includegraphics[width=\linewidth]{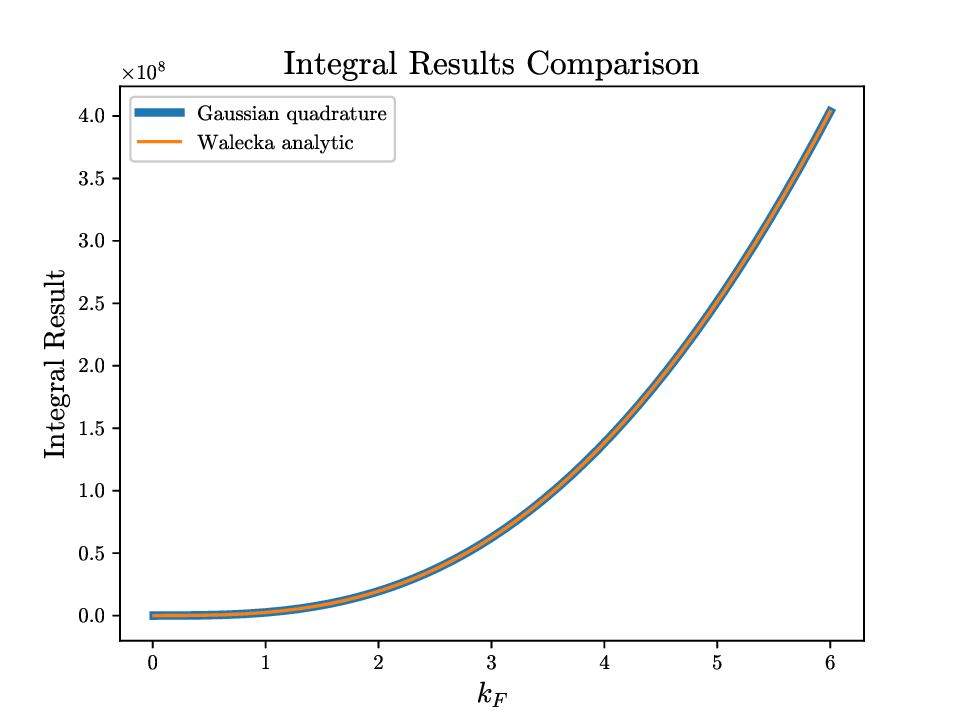}
	\caption{Validation of the scalar density integral: numerical vs. analytic evaluation.}
	\label{gb4.1}
\end{figure}

The scalar density integral (equation (\ref{eq:1.3})) was re-derived in detail (Appendix \ref{secA1}) and evaluated numerically using a 20-point Gaussian quadrature. As shown in Figure \ref{gb4.1}, the numerical and analytic results overlap almost perfectly, confirming both the correctness of Walecka’s original derivation \citep{wala} and the reliability of the numerical routine. Establishing this baseline is essential, since extensions of QHD-I with additional meson fields will rely on the same numerical framework.

\subsection{Effective Mass and Parameter Set Dependence}

\begin{figure}[t]
	\centering
	\includegraphics[width=\linewidth]{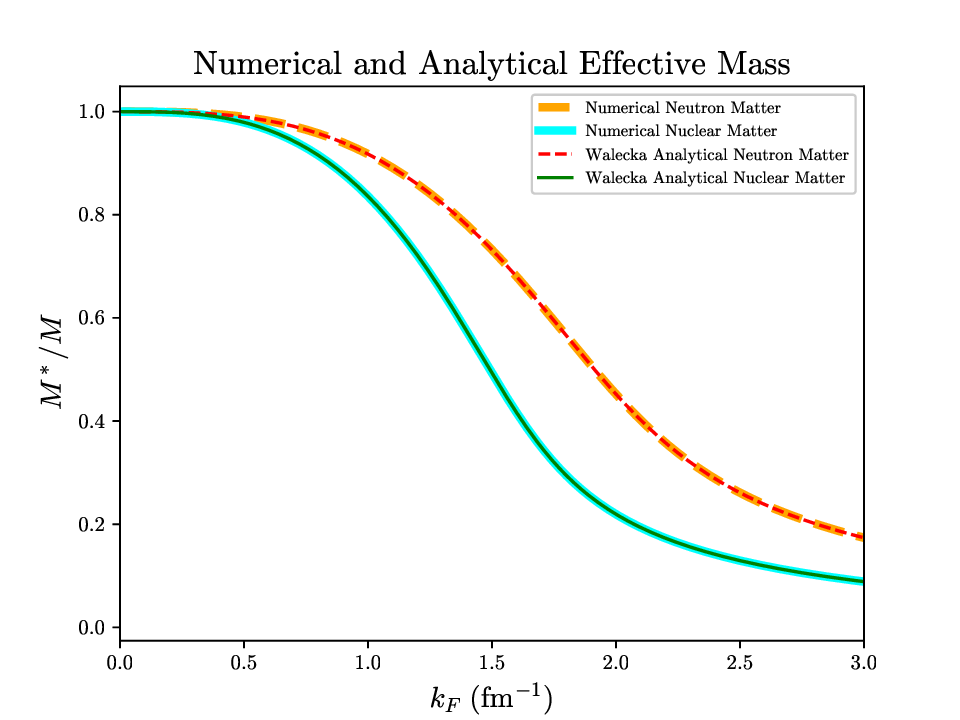}
	\caption{Relative effective mass $M^*/M$ from numerical and analytic evaluation.}
	\label{gb4.2}
\end{figure}

The relative effective mass was computed numerically and analytically, with results in Figure \ref{gb4.2} showing near-perfect overlap. This validates the self-consistent solution method and highlights $M^*$ as a robust output of the model.

Physically, the decrease of $M^*$ with density reflects the competition between a strong scalar attraction ($g_s \phi_0 \cong 400$ MeV) and a comparably strong vector repulsion ($g_v V_0 \cong 330$ MeV), whose near cancellation yields the small nuclear binding energy. This Lorentz-structured mechanism, absent in nonrelativistic models, drives nuclear saturation \citep{swa}.

\begin{figure}[t]
	\centering
	\includegraphics[width=\linewidth]{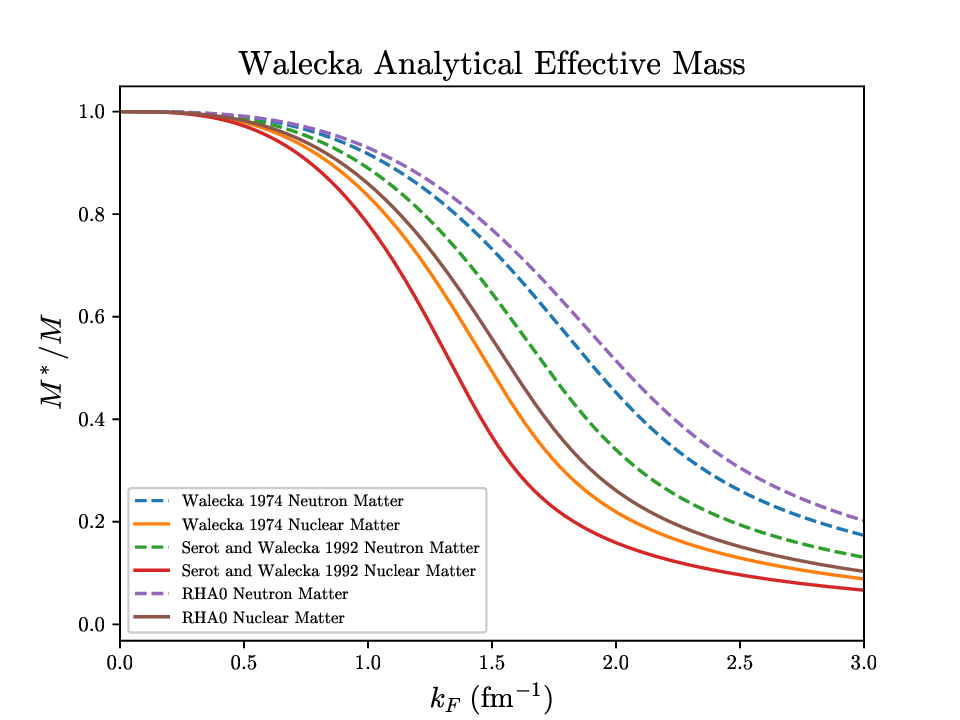}
	\caption{Effective mass for different parameter sets.}
	\label{gb4.3}
\end{figure}

Parameter dependence is illustrated in Figure \ref{gb4.3}. In the original work by \citet{wala}, the two coupling constants $\{C_s^2, C_v^2\}$ were explicitly fitted to reproduce the empirical binding energy per nucleon of $-15.75$ MeV at $k_F = 1.42$ fm$^{-1}$. Once these two parameters are fixed, all other quantities in the model are predictions. This procedure leads to notable consequences: the binding energy curve develops a characteristic dip near nuclear densities, neutron matter appears unbound (in agreement with other theoretical calculations), and later the resulting EoS is relatively stiff at high densities. Thus, the act of parameter fitting not only determines saturation properties but also propagates into qualitative features of the EoS.

\begin{figure}[t]
	\centering
	\includegraphics[width=\linewidth]{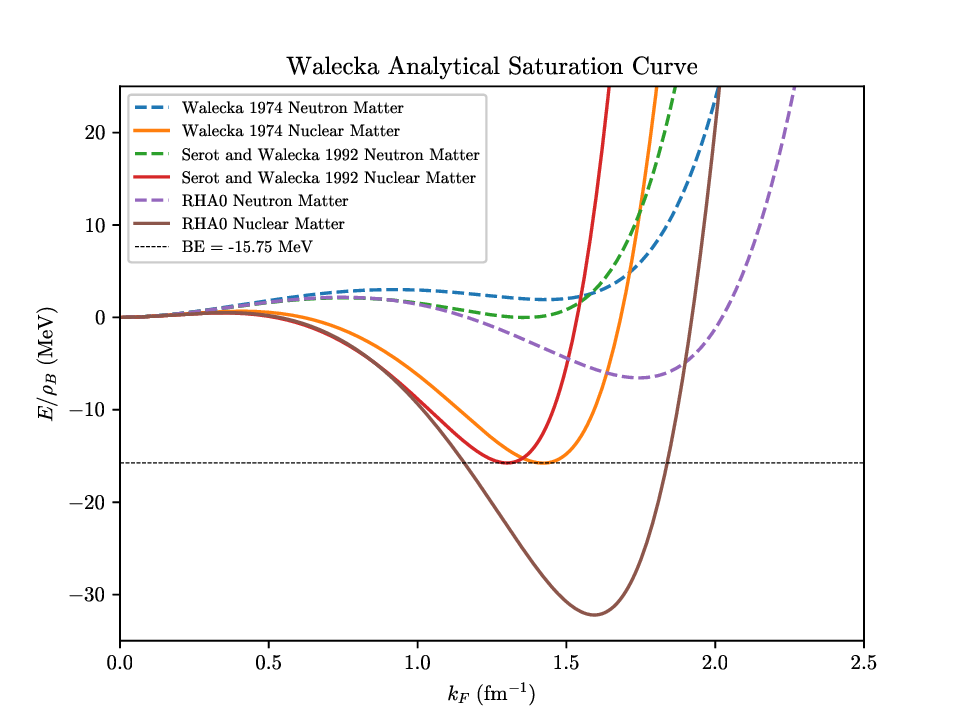}
	\caption{Binding energy per nucleon for parameter sets comparison.}
	\label{gb4.4}
\end{figure}

Binding energy curves for different parameterizations (Figure \ref{gb4.4}) further demonstrate how the choice of couplings influences compressibility and saturation. For instance, the \citet{swa} set shifts the saturation point to $k_F=1.30$ fm$^{-1}$ with a similar binding depth, while the RHA0 set derived from a Dirac–Hartree treatment produces deeper binding. These differences emphasize that parameter sets are not unique; they reflect distinct fitting strategies, either to empirical observables or to theoretical benchmarks.

Beyond QHD-I, comparative studies by Diener \cite{die} examined a wide range of relativistic mean-field parameterizations—such as QHD-I, NL3 \citep{lalazissis}, PK1 \citep{long}, and FSUGold \citep{piekarewicz}—which extend the original model with nonlinear scalar interactions, additional mesons (e.g. $\rho$), or modified couplings. Importantly, these advanced sets are also fitted to different inputs: some to nuclear saturation properties or finite nuclei data like charge radii. Below saturation ($k_F\sim1.3$ fm$^{-1}$), all such parameterizations yield almost identical binding energies, but their predictions diverge significantly at higher densities. This highlights both the common foundation of QHD-based models and the sensitivity of dense matter predictions to the details of parameter fitting.

\subsection{Equation of State Analysis}

\begin{figure}[t]
	\centering
	\includegraphics[width=\linewidth]{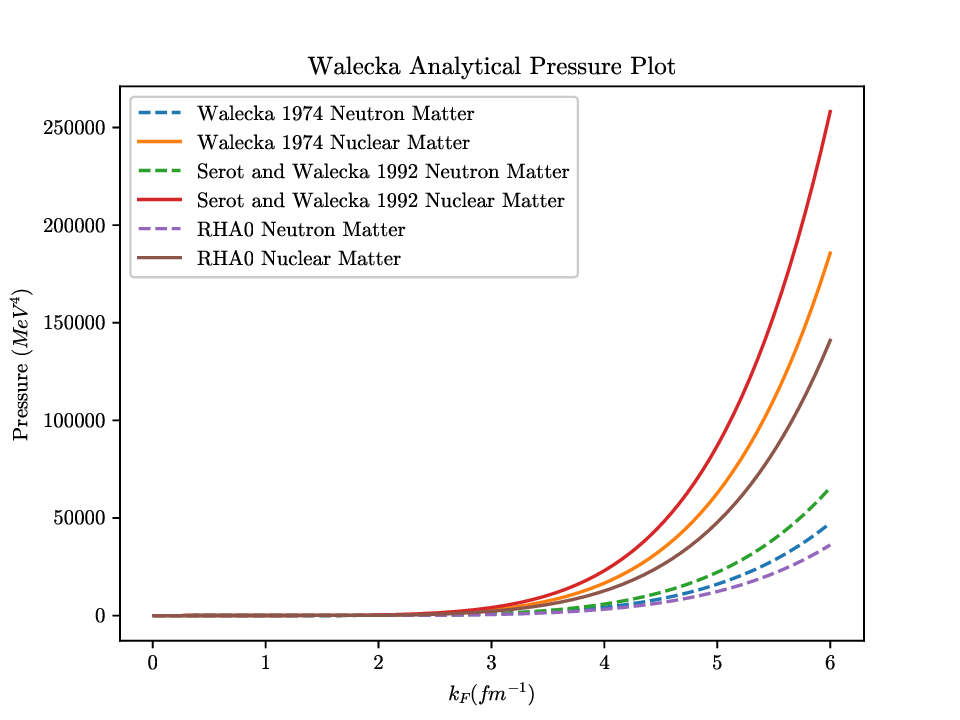}
	\caption{Pressure vs. Fermi momentum for three parameter sets.}
	\label{gb4.13}
\end{figure}

The pressure as a function of Fermi momentum (Figure \ref{gb4.13}) exhibits the expected monotonic increase, with slopes differing across parameter sets. The Serot and Walecka \cite{swa} set yields the steepest rise, while RHA0 is more gradual, illustrating how parameterization controls EoS stiffness.

\begin{figure}[t]
	\centering
	\includegraphics[width=\linewidth]{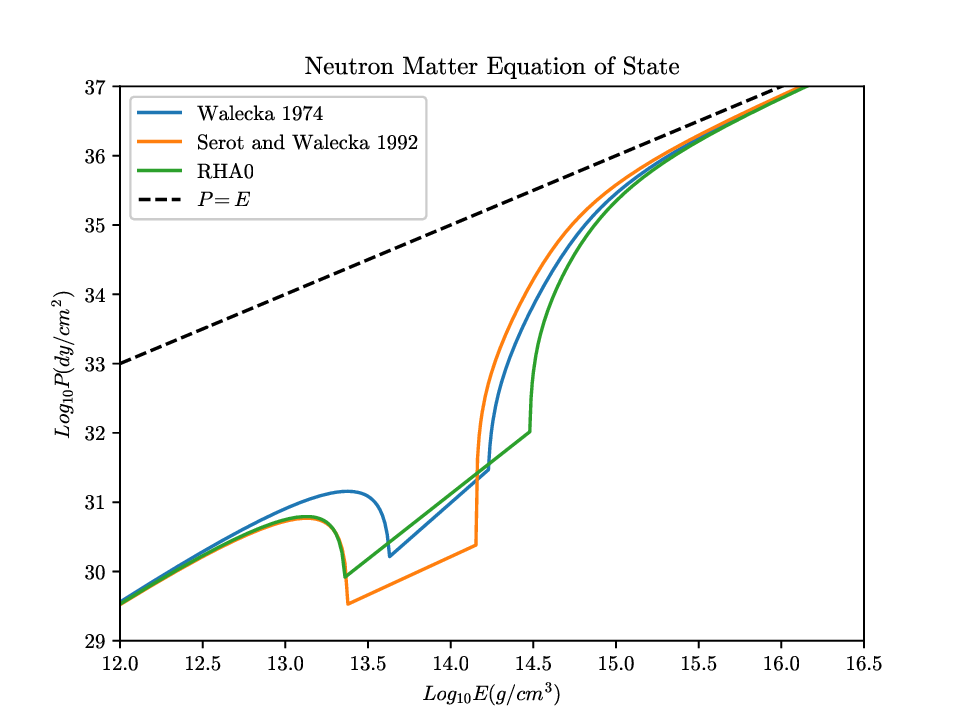}
	\caption{Equation of state: logarithmic pressure vs. energy density for parameter sets.}
	\label{gb4.14}
\end{figure}

The logarithmic pressure–energy density relation (Figure \ref{gb4.14}) spans $\sim 10^{13.4}$–$10^{14.5}$ g/cm$^3$ and approaches the causal limit ($p=\varepsilon$) at high density. As Serot and Walecka \cite{swa} noted, the QHD-I mean-field model resembles a van der Waals system with a liquid–gas transition and causal extrapolation to high densities, inherently producing a stiff EoS. While stiff EoS maximize neutron star mass, they are unrealistic without softening mechanisms such as $\beta$-equilibrium and the inclusion of additional particle species (e.g. protons, electrons, muons) \citep{die}.

Formally, the QHD-I EoS is given by
\begin{equation}
	\varrho_B = \frac{\gamma}{2\pi^2}\int_0^{k_F} dk=\frac{\gamma}{6\pi^2} k_F^3,
\end{equation}
\begin{align}
	\mathcal{E} = \frac{g_v^2}{2m_v^2} \varrho_B^2 + \frac{m_s^2}{2g_s^2}(M-M^*)^2 \notag \\
	+ \frac{\gamma}{2\pi^2}\int_0^{k_F} dk \,k^2\sqrt{k^2+M^{*2}},
\end{align}
\begin{align}
	p = \frac{g_v^2}{2m_v^2} \varrho_B^2 - \frac{m_s^2}{2g_s^2}(M-M^*)^2 \notag \\
	+ \frac{\gamma}{6\pi^2}\int_0^{k_F} dk \,\frac{k^4}{\sqrt{k^2+M^{*2}}},
\end{align}
where $\varrho_B$ is the baryon density (here equivalent to nucleon density). Here $M^*$ enters explicitly in two of the three terms, confirming its central role in connecting meson–nucleon couplings to macroscopic observables. Parameter shifts propagate through $M^*$ to alter pressure and energy density, thereby controlling EoS stiffness.

Overall, QHD-I provides a consistent baseline: it captures qualitative features, establishes the causal limit, and demonstrates the microscopic–macroscopic link through $M^*$. Yet, its intrinsic simplicity yields overly stiff EoS. Realistic neutron star modeling requires extensions—such as nonlinear scalar terms, $\rho$-meson interactions, or additional particle degrees of freedom—to soften the EoS and better match astrophysical constraints.

\section{Conclusions}

This work revisited the effective mass formulation in the QHD-I (Walecka) model through a combined analytic and numerical analysis. By rederiving the scalar density integral in detail and validating it with Gaussian quadrature, we confirmed that the self-consistent effective mass relation of \citet{wala} is correct and numerically robust. This provides a solid methodological basis for extending such calculations to more complex hadronic models.

The comparative study of parameter sets demonstrated that variations in meson masses and couplings strongly affect the effective mass, binding energy, and compressibility. Nevertheless, within the QHD-I framework, all parameter choices converge to a relatively stiff equation of state. This stiffness arises from the restricted degrees of freedom—nucleons, a scalar meson, and a vector meson—and highlights the intrinsic limitations of the model for realistic neutron star matter.

Despite these limitations, QHD-I remains an instructive baseline that highlights the central role of the effective mass $M^*$ as a microscopic–macroscopic link: scalar attraction and vector repulsion at the nucleon level translate directly into bulk saturation and stiffness at the neutron-star scale. The numerical program established in this work is not limited to QHD-I; with minimal modification it can be applied to RMF models that incorporate nonlinear couplings, additional meson fields, or $\beta$-equilibrated matter. Such extensions are essential for obtaining softer and more realistic EoS and for connecting theoretical nuclear models to astrophysical observations of neutron stars, including maximum masses, radii, and tidal deformabilities.

\bmhead{Acknowledgements}

The initial version of this study was conducted as part of an undergraduate thesis. The authors confirm that this manuscript contains original material not previously published or submitted elsewhere.

\section*{Declarations}

%Some journals require declarations to be submitted in a standardised format. Please check the Instructions for Authors of the journal to which you are submitting to see if you need to complete this section. If yes, your manuscript must contain the following sections under the heading `Declarations':

\begin{itemize}
	\item Funding:
	No funding was received for conducting this study.
	\item Competing interests:
	The authors have no relevant financial or non-financial interests to disclose.
%	\item Ethics approval and consent to participate
%	\item Consent for publication
%	\item Data availability 
%	\item Materials availability
	\item Code availability:
	The source code developed and used in this study is openly available at \url{https://github.com/ghitha-rahiemy/Modified-Numerical-QHD-effective-mass}. Additional data generated or analyzed during the current study are available from the corresponding author on reasonable request.
	\item Author contribution:
	All authors contributed to the study conception and design. Material preparation, data collection, analysis, and manuscript writing were performed by Ghitha Nadhira Azka Rahiemy. Eko Tri Sulistyani provided guidance on neutron star physics and interpretation of astrophysical aspects. Pekik Nurwantoro contributed to the numerical methods, program development guidance, and provided computational facilities. All authors reviewed and approved the final version of the manuscript.
\end{itemize}

\begin{appendices}

\section{Scalar Density Integral Derivation}\label{secA1}

This section presents the derivation of the scalar density integral used in this study:
\begin{equation}
    \int^{k_F}_0 dk \frac{k^2 M^*}{\sqrt{k^2+M^*}} = M^* \int_0^{k_F} \frac{k^2}{\sqrt{k^2+M^{*2}}}
\end{equation}
The integral without effective mass (a constant) is defined:
\begin{equation}
	\int_0^{k_F} \frac{k^2}{\sqrt{k^2+M^{*2}}} = I
\end{equation}
Using integration by parts, let us define $k = x$ and $M^* = a$, with
\begin{align}
    u &= x \\
    dv &= dx \frac{x}{\sqrt{x^2+a^2}}
\end{align}
Differentiating these expressions gives:
\begin{align}
    du &= dx \\
    v &= \int dx \frac{x}{\sqrt{x^2+a^2}}=\sqrt{x^2+a^2}+C_v \label{eq:B.3b}
\end{align}
The result of Equation (\ref{eq:B.3b}) is derived using the following relation:
\begin{equation}
    \frac{d(\sqrt{x^2+a^2})}{dx}=\frac{x}{\sqrt{x^2+a^2}} 
\end{equation}
Substituting these expressions into the general formula for integration by parts, we obtain:
\begin{equation}
    uv - \int v \, du = x \sqrt{x^2+a^2} - \int \sqrt{x^2+a^2} \, dx
      \label{eq:B.5}
\end{equation}
The second integral in Equation (\ref{eq:B.5}) is further solved by applying integration by parts again, where:
\begin{align}
    u &= \sqrt{x^2+a^2} \\
    dv &= dx 
\end{align}
Differentiating these gives:
\begin{align}
    du &= \frac{x}{\sqrt{x^2+a^2}} \, dx \\
    v &= x
\end{align}
Substituting these variables back, we find:
\begin{align}
    \int \sqrt{x^2 + a^2} \, dx &= x \sqrt{x^2 + a^2} \notag \\
    - \int \frac{x^2}{\sqrt{x^2 + a^2}} \, dx = S  \\
    S &= x \sqrt{x^2 + a^2} - I 
\end{align}
To evaluate the integral \(I\), we rewrite it as:
\begin{align}
    \int \frac{x^2}{\sqrt{x^2 + a^2}} \, dx &= \int \left( \sqrt{x^2 + a^2} - \frac{a^2}{\sqrt{x^2 + a^2}} \right) \, dx \\
    I &= S - a^2 J  \label{eq:B.11}
\end{align}
where
\begin{equation}
    J = \int \frac{1}{\sqrt{x^2 + a^2}} \, dx 
\end{equation}
Using the substitution:
\begin{align}
    x &= a \sinh \theta  \\
    dx &= a \cosh \theta \, d\theta  \\
    \sqrt{x^2 + a^2} &= a \cosh \theta 
\end{align}
Substituting these variables into \(J\), we get:
\begin{align}
    J &= \int d\theta = \theta + C 
\end{align}
Since:
\begin{equation}
    \theta = \sinh^{-1} \left( \frac{x}{a} \right) = \ln \left( \frac{x + \sqrt{x^2 + a^2}}{a} \right) 
\end{equation}
Thus,
\begin{equation}
    J = \ln \left( \frac{x + \sqrt{x^2 + a^2}}{a} \right) + C  \label{eq:B.16}
\end{equation}
Using Equation (\ref{eq:B.16}) in (\ref{eq:B.11}), we obtain:
\begin{align}
    I &= S - a^2 J \\
    2I &= x \sqrt{x^2 + a^2} - a^2 \ln \left( \frac{x + \sqrt{x^2 + a^2}}{a} \right) \\
    I &= \frac{1}{2} \left[ x \sqrt{x^2 + a^2} - a^2 \ln \left( \frac{x + \sqrt{x^2 + a^2}}{a} \right) \right] 
\end{align}
Finally, substituting the original variables, we obtain the expression:
\begin{align}
    \int^{k_F}_0 dk \frac{k^2 M^*}{\sqrt{k^2 + M^{*2}}} = \frac{M^*}{2} \left[ k_F \sqrt{k_F^2 + M^{*2}} \right. \notag \\
    - M^{*2} \ln \left. \left( \frac{k_F + \sqrt{k_F^2 + M^{*2}}}{M^*} \right) \right] 
\end{align}




\end{appendices}


\bibliography{sn-bibliography}% common bib file


\begin{thebibliography}{19}
\ifx \bisbn   \undefined \def \bisbn  #1{ISBN #1}\fi
\ifx \binits  \undefined \def \binits#1{#1}\fi
\ifx \bauthor  \undefined \def \bauthor#1{#1}\fi
\ifx \batitle  \undefined \def \batitle#1{#1}\fi
\ifx \bjtitle  \undefined \def \bjtitle#1{#1}\fi
\ifx \bvolume  \undefined \def \bvolume#1{\textbf{#1}}\fi
\ifx \byear  \undefined \def \byear#1{#1}\fi
\ifx \bissue  \undefined \def \bissue#1{#1}\fi
\ifx \bfpage  \undefined \def \bfpage#1{#1}\fi
\ifx \blpage  \undefined \def \blpage #1{#1}\fi
\ifx \burl  \undefined \def \burl#1{\textsf{#1}}\fi
\ifx \doiurl  \undefined \def \doiurl#1{\url{https://doi.org/#1}}\fi
\ifx \betal  \undefined \def \betal{\textit{et al.}}\fi
\ifx \binstitute  \undefined \def \binstitute#1{#1}\fi
\ifx \binstitutionaled  \undefined \def \binstitutionaled#1{#1}\fi
\ifx \bctitle  \undefined \def \bctitle#1{#1}\fi
\ifx \beditor  \undefined \def \beditor#1{#1}\fi
\ifx \bpublisher  \undefined \def \bpublisher#1{#1}\fi
\ifx \bbtitle  \undefined \def \bbtitle#1{#1}\fi
\ifx \bedition  \undefined \def \bedition#1{#1}\fi
\ifx \bseriesno  \undefined \def \bseriesno#1{#1}\fi
\ifx \blocation  \undefined \def \blocation#1{#1}\fi
\ifx \bsertitle  \undefined \def \bsertitle#1{#1}\fi
\ifx \bsnm \undefined \def \bsnm#1{#1}\fi
\ifx \bsuffix \undefined \def \bsuffix#1{#1}\fi
\ifx \bparticle \undefined \def \bparticle#1{#1}\fi
\ifx \barticle \undefined \def \barticle#1{#1}\fi
\bibcommenthead
\ifx \bconfdate \undefined \def \bconfdate #1{#1}\fi
\ifx \botherref \undefined \def \botherref #1{#1}\fi
\ifx \url \undefined \def \url#1{\textsf{#1}}\fi
\ifx \bchapter \undefined \def \bchapter#1{#1}\fi
\ifx \bbook \undefined \def \bbook#1{#1}\fi
\ifx \bcomment \undefined \def \bcomment#1{#1}\fi
\ifx \oauthor \undefined \def \oauthor#1{#1}\fi
\ifx \citeauthoryear \undefined \def \citeauthoryear#1{#1}\fi
\ifx \endbibitem  \undefined \def \endbibitem {}\fi
\ifx \bconflocation  \undefined \def \bconflocation#1{#1}\fi
\ifx \arxivurl  \undefined \def \arxivurl#1{\textsf{#1}}\fi
\csname PreBibitemsHook\endcsname

%%% 1
\bibitem[\protect\citeauthoryear{Antic and et~al.}{2018}]{antic}
\begin{botherref}
\oauthor{\bsnm{Antic}, \binits{S.}},
\oauthor{\bsnm{al.}}:
Generalized relativistic mean-field model with non-linear derivative
  nucleon-meson couplings for nuclear matter and finite nuclei.
Technical report,
Technische Univ. Darmstadt (Germany). Fachbereich Physik
(2018)
\end{botherref}
\endbibitem

%%% 2
\bibitem[\protect\citeauthoryear{Diener}{2008}]{die}
\begin{botherref}
\oauthor{\bsnm{Diener}, \binits{J.P.W.}}:
Relativistic mean-field theory applied to the study of neutron star properties.
PhD thesis,
Stellenbosch University
(2008).
\url{https://doi.org/10.48550/ARXIV.0806.0747}
\end{botherref}
\endbibitem

%%% 3
\bibitem[\protect\citeauthoryear{Frishman and Sonnenschein}{2010}]{Frishman}
\begin{bbook}
\bauthor{\bsnm{Frishman}, \binits{Y.}},
\bauthor{\bsnm{Sonnenschein}, \binits{J.}}:
\bbtitle{Non-Perturbative Field Theory: From Two Dimensional Conformal Field
  Theory to QCD in Four Dimensions}.
\bsertitle{Cambridge Monographs on Mathematical Physics}.
\bpublisher{Cambridge University Press},
\blocation{Cambridge, UK}
(\byear{2010})
\end{bbook}
\endbibitem

%%% 4
\bibitem[\protect\citeauthoryear{Dupuis et~al.}{2021}]{dupuis}
\begin{barticle}
\bauthor{\bsnm{Dupuis}, \binits{N.}},
\bauthor{\bsnm{Canet}, \binits{L.}},
\bauthor{\bsnm{Eichhorn}, \binits{A.}},
\bauthor{\bsnm{Metzner}, \binits{W.}},
\bauthor{\bsnm{Pawlowski}, \binits{J.M.}},
\bauthor{\bsnm{Tissier}, \binits{M.}},
\bauthor{\bsnm{Wschebor}, \binits{N.}}:
\batitle{The nonperturbative functional renormalization group and its
  applications}.
\bjtitle{Phys. Rep.}
\bvolume{910},
\bfpage{1}--\blpage{114}
(\byear{2021})
\doiurl{10.1016/j.physrep.2021.01.001}
\end{barticle}
\endbibitem

%%% 5
\bibitem[\protect\citeauthoryear{Abbott et~al.}{2025}]{abbott}
\begin{barticle}
\bauthor{\bsnm{Abbott}, \binits{R.}},
\bauthor{\bsnm{Detmold}, \binits{W.}},
\bauthor{\bsnm{Illa}, \binits{M.}},
\bauthor{\bsnm{Parreño}, \binits{A.}},
\bauthor{\bsnm{Perry}, \binits{R.J.}},
\bauthor{\bsnm{Romero-López}, \binits{F.}},
\bauthor{\bsnm{Shanahan}, \binits{P.E.}},
\bauthor{\bsnm{Wagman}, \binits{M.L.}}:
\batitle{Qcd constraints on isospin-dense matter and the nuclear equation of
  state}.
\bjtitle{Phys. Rev. Lett.}
\bvolume{134}(\bissue{1}),
\bfpage{011903}
(\byear{2025})
\doiurl{10.1103/PhysRevLett.134.011903}
\end{barticle}
\endbibitem

%%% 6
\bibitem[\protect\citeauthoryear{Kumar et~al.}{2025}]{kumar}
\begin{botherref}
\oauthor{\bsnm{Kumar}, \binits{R.}},
\oauthor{\bsnm{Dexheimer}, \binits{V.}},
\oauthor{\bsnm{Jahan}, \binits{J.}}:
Neutron stars and Constraints for the Equation of State of Dense Matter.
arXiv preprint arXiv:2503.23413
(2025).
\doiurl{10.48550/ARXIV.2503.23413}
\end{botherref}
\endbibitem

%%% 7
\bibitem[\protect\citeauthoryear{Walecka}{1974}]{wala}
\begin{barticle}
\bauthor{\bsnm{Walecka}, \binits{J.D.}}:
\batitle{A theory of highly condensed matter}.
\bjtitle{Ann. Phys.}
\bvolume{83},
\bfpage{491}--\blpage{529}
(\byear{1974})
\doiurl{10.1016/0003-4916(74)90208-5}
\end{barticle}
\endbibitem

%%% 8
\bibitem[\protect\citeauthoryear{Negele et~al.}{1986}]{sw}
\begin{bbook}
\bauthor{\bsnm{Negele}, \binits{J.W.}},
\bauthor{\bsnm{Serot}, \binits{B.D.}},
\bauthor{\bsnm{Vogt}, \binits{E.}},
\bauthor{\bsnm{Walecka}, \binits{J.D.}}:
\bbtitle{The Relativistic Nuclear Many-body Problem}.
\bpublisher{Plenum Press},
\blocation{New York, NY}
(\byear{1986}).
\burl{https://www.osti.gov/biblio/5443763}
\end{bbook}
\endbibitem

%%% 9
\bibitem[\protect\citeauthoryear{Erkelenz}{1974}]{er}
\begin{barticle}
\bauthor{\bsnm{Erkelenz}, \binits{K.}}:
\batitle{Current status of the relativistic two-nucleon one boson exchange
  potential}.
\bjtitle{Phys. Rep.}
\bvolume{13},
\bfpage{191}--\blpage{258}
(\byear{1974})
\doiurl{10.1016/0370-1573(74)90008-8}
\end{barticle}
\endbibitem

%%% 10
\bibitem[\protect\citeauthoryear{Holinde}{1981}]{ho}
\begin{barticle}
\bauthor{\bsnm{Holinde}, \binits{K.}}:
\batitle{Two-nucleon forces and nuclear matter}.
\bjtitle{Phys. Rep.}
\bvolume{68},
\bfpage{121}--\blpage{188}
(\byear{1981})
\doiurl{10.1016/0370-1573(81)90188-5}
\end{barticle}
\endbibitem

%%% 11
\bibitem[\protect\citeauthoryear{Bednarek}{2007}]{bednarek}
\begin{bbook}
\bauthor{\bsnm{Bednarek}, \binits{I.}}:
\bbtitle{Relativistic Mean Field: Models of Neutron Stars}.
\bpublisher{Wydawnictwo Uniwersytetu Śląskiego},
\blocation{Katowice, Poland}
(\byear{2007})
\end{bbook}
\endbibitem

%%% 12
\bibitem[\protect\citeauthoryear{Huang et~al.}{2024}]{huang}
\begin{barticle}
\bauthor{\bsnm{Huang}, \binits{C.}}, \betal:
\batitle{Constraining a relativistic mean field model using neutron star
  mass–radius measurements i: nucleonic models}.
\bjtitle{Mon. Not. R. Astron. Soc.}
\bvolume{529}(\bissue{4}),
\bfpage{4650}--\blpage{4665}
(\byear{2024})
\doiurl{10.1093/mnras/stae844}
\end{barticle}
\endbibitem

%%% 13
\bibitem[\protect\citeauthoryear{Pais et~al.}{2022}]{pais}
\begin{barticle}
\bauthor{\bsnm{Pais}, \binits{H.}},
\bauthor{\bsnm{Ivanytskyi}, \binits{O.}},
\bauthor{\bsnm{Providência}, \binits{C.}}:
\batitle{Landau parameters and entrainment matrix of cold stellar matter:
  effect of the symmetry energy and strong magnetic fields}.
\bjtitle{J. Cosmol. Astropart. Phys.}
\bvolume{2022}(\bissue{04}),
\bfpage{024}
(\byear{2022})
\doiurl{10.1088/1475-7516/2022/04/024}
\end{barticle}
\endbibitem

%%% 14
\bibitem[\protect\citeauthoryear{Scurto et~al.}{2024}]{scurto}
\begin{botherref}
\oauthor{\bsnm{Scurto}, \binits{L.}},
\oauthor{\bsnm{Carvalho}, \binits{V.}},
\oauthor{\bsnm{Pais}, \binits{H.}},
\oauthor{\bsnm{Providência}, \binits{C.}}:
Assessing the joint effect of temperature and magnetic field on the neutron
  star equation of state.
arXiv preprint arXiv:2407.03113
(2024).
\doiurl{10.48550/ARXIV.2407.03113}
\end{botherref}
\endbibitem

%%% 15
\bibitem[\protect\citeauthoryear{Wang et~al.}{2022}]{wang}
\begin{barticle}
\bauthor{\bsnm{Wang}, \binits{X.}},
\bauthor{\bsnm{Li}, \binits{J.}},
\bauthor{\bsnm{Fang}, \binits{J.}},
\bauthor{\bsnm{Pais}, \binits{H.}},
\bauthor{\bsnm{Providência}, \binits{C.}}:
\batitle{Pasta phases in neutron stars under strong magnetic fields}.
\bjtitle{Phys. Rev. D}
\bvolume{105}(\bissue{6}),
\bfpage{063004}
(\byear{2022})
\doiurl{10.1103/PhysRevD.105.063004}
\end{barticle}
\endbibitem

%%% 16
\bibitem[\protect\citeauthoryear{Serot and Walecka}{1992}]{swa}
\begin{bchapter}
\bauthor{\bsnm{Serot}, \binits{B.D.}},
\bauthor{\bsnm{Walecka}, \binits{J.D.}}:
\bctitle{Relativistic nuclear many-body theory}.
In: \beditor{\bsnm{Ainsworth}, \binits{T.L.}},
\beditor{\bsnm{Campbell}, \binits{C.E.}},
\beditor{\bsnm{Clements}, \binits{B.E.}},
\beditor{\bsnm{Krotscheck}, \binits{E.}} (eds.)
\bbtitle{Recent Progress in Many-Body Theories: Volume 3},
pp. \bfpage{49}--\blpage{92}.
\bpublisher{Springer},
\blocation{Boston, MA}
(\byear{1992}).
\doiurl{10.1007/978-1-4615-3466-2_5}
\end{bchapter}
\endbibitem

%%% 17
\bibitem[\protect\citeauthoryear{Lalazissis et~al.}{1997}]{lalazissis}
\begin{barticle}
\bauthor{\bsnm{Lalazissis}, \binits{G.A.}},
\bauthor{\bsnm{König}, \binits{J.}},
\bauthor{\bsnm{Ring}, \binits{P.}}:
\batitle{New parametrization for the lagrangian density of relativistic mean
  field theory}.
\bjtitle{Phys. Rev. C}
\bvolume{55}(\bissue{1}),
\bfpage{540}--\blpage{543}
(\byear{1997})
\doiurl{10.1103/PhysRevC.55.540}
\end{barticle}
\endbibitem

%%% 18
\bibitem[\protect\citeauthoryear{Long et~al.}{2004}]{long}
\begin{barticle}
\bauthor{\bsnm{Long}, \binits{W.}},
\bauthor{\bsnm{Meng}, \binits{J.}},
\bauthor{\bsnm{Giai}, \binits{N.V.}},
\bauthor{\bsnm{Zhou}, \binits{S.-G.}}:
\batitle{New effective interactions in relativistic mean field theory with
  nonlinear terms and density-dependent meson-nucleon coupling}.
\bjtitle{Phys. Rev. C}
\bvolume{69}(\bissue{3}),
\bfpage{034319}
(\byear{2004})
\doiurl{10.1103/PhysRevC.69.034319}
\end{barticle}
\endbibitem

%%% 19
\bibitem[\protect\citeauthoryear{Piekarewicz}{2007}]{piekarewicz}
\begin{barticle}
\bauthor{\bsnm{Piekarewicz}, \binits{J.}}:
\batitle{Validating relativistic models of nuclear structure against
  theoretical, experimental, and observational constraints}.
\bjtitle{Phys. Rev. C}
\bvolume{76}(\bissue{6}),
\bfpage{064310}
(\byear{2007})
\doiurl{10.1103/PhysRevC.76.064310}
\end{barticle}
\endbibitem

\end{thebibliography}

\end{document}